\documentstyle[preprint,aps]{revtex}
\begin{document}
\draft
%\preprint{\vbox{\hfill TCTS9502\\}}
%
\title{
Scale Anomaly and Quantum Chaos
in the Billiards with Pointlike Scatterers
}
\author{
{Taksu Cheon}
}
\address{
Department of Physics, Hosei University, Fujimi, 
Chiyoda-ku, Tokyo 102, Japan
}
\author{
{Takaomi Shigehara}
}
\address{
Computer Centre, University of Tokyo, Yayoi, 
Bunkyo-ku, Tokyo 113, Japan 
}
%
%\date{\today}
\date{October 3, 1995}
\maketitle
\begin{abstract}
We argue that the random-matrix like energy spectra found in 
pseudointegrable billiards with pointlike scatterers are 
related to the quantum violation of scale invariance of 
classical analogue system.  It is shown that the behavior 
of the running coupling constant explains the key 
characteristics of the level statistics of pseudointegrable 
billiards.
\end{abstract}
\pacs{5.45.+b, 3.65.Db, 11.10.Gh}
%
%\vfill{\ \\
%PACS Nos: 3.65.Db, 5.45.+b, 11.10.Gh\\
%-----------------------------------------------------\\
%email contact: cheon@fujimi.hosei.ac.jp, takaomi@cc.u-tokyo.ac.jp}
%
%\narrowtext
%\twocolumn
%\newpage
\paragraph*{}
The concepts of scale anomaly and the asymptotic freedom are among 
the key features of the gauge field theories which describe the 
interaction of the elementary particles.  It is less widely 
recognized, however, that the scale anomaly can be found in 
a vastly simpler setting of one particle quantum mechanics.  
A particle scattered off a pointlike scatterer 
in two spacial dimension is known to have energy dependent 
$s$-wave phase shift 
defying the scale invariance of its classical analogue \cite{JA95}.  
There exists a sister problem of particle motion  confined in 
a hard-wall boundary with a pointlike scatterer inside.  
When the shape of the boundary is a rectangle, the problem belongs 
to a larger category of systems known as pseudointegrable 
billiards \cite{RB81,CC89,SE90,SH94}.  
This system is known for puzzling statistical 
properties of its energy eigenvalues \cite{CC89,SH94}.  
It is shown through numerical experiments that the level 
statistics of the pseudointegrable billiard resembles to that of 
random-matrix ensembles \cite{ME90} which is generally associated 
with chaotic dynamics \cite{BG83}.  
This is in seeming contradiction with  
the absence of chaotic dynamics in classical analogue system.  
Further, when the levels are collected at higher energy region,
the level statistics moves toward the Poisson distribution which 
characterizes the integrable classical dynamics.  Also, the 
system tends to show more of Wigner-like statistics when the 
genus of the billiard is increased, that is, in the present 
context, when the number of the singular scatterers is increased.  
These facts have never received sufficient explanations, 
in spite of several attempted studies based on the semiclassical 
periodic orbit quantization theory \cite{BS93,SS93}.

        In this paper, we argue that the behavior of spectra of 
the pseudointegrable billiard with pointlike scatterers is a 
direct result of the scale anomaly of the system.  Specifically, 
the dependence of the level statistics on the energy and the 
number of the scatterers is shown to be well explained by the 
high energy behavior of the effective coupling strength of 
the pointlike scatterer.

        We consider a quantum particle of unit mass moving 
freely inside a boundary $B$ in two spacial dimension on 
which its wave functions are assumed to vanish.  We denote 
the eigenvalue and eigenfunction of the system as
$\varepsilon _n$ and $\phi _n$, namely
\begin{mathletters}
\label{1}
\begin{eqnarray}
\label{1a}
-{1 \over 2}\nabla ^2\ \phi _n({\vec x})
=\varepsilon _n\ \phi _n({\vec x})
\end{eqnarray}
with
\begin{eqnarray}
\label{1b}
 \phi _n({\vec x_B})=0
\ \ \   {\rm where} \ \ \  
{\vec x_B}\in B.
\end{eqnarray}
\end{mathletters}
Assuming $\phi_n({\vec x})$ to be normalized to unity, the Green's 
function is given by
\begin{eqnarray}
\label{2}
G^{(0)}({\vec x},{\vec x'};\omega )
=\sum\limits_{n=1}^\infty 
{{{\phi _n({\vec x})\phi _n({\vec x'})} 
\over {\omega -\varepsilon _n}}}.
\end{eqnarray}
When the shape of the boundary $B$ is such that the classical motion
of the particle is regular (as in the case of $B$ being rectangular),
the nearest neighbor spacing $s_n=\varepsilon_{n+1}-\varepsilon_n$
 is distributed according to the
Poisson distribution $P(s)=\exp (-s)$.   
We now place a pointlike scatterer at ${\vec x_0}$.
Naively, one defines the scatterer in terms of the 
Dirac's delta function in two dimension
\begin{eqnarray}
\label{3}
V({\vec x})=v\ \delta ({\vec x}-{\vec x_0}).
\end{eqnarray}
Under the scale transformation ${\vec x}\to a{\vec x}$, 
the potential is 
transformed as $V({\vec x})\to {V({\vec x})}/{a^2}$.
Since this behavior is identical to that of the Laplacian operator 
in eq. (\ref{1a}), the system is {\em scale invariant}.  
One expects, therefore, 
that the dynamical properties of the system should not depend on 
the energy.  Formally, the transition matrix $T$ (T-matrix) in the 
presence of the scatterer $V$ is given by
\begin{eqnarray}
\label{4}
T=\ V+\ V\ G^{(0)}T.
\end{eqnarray}
The poles of $T$ give the eigenvalues of the system.  
Because of the separability of the delta potential,
$\langle \phi _n\ V\ \phi _m\rangle
=v\ \phi _n({\vec x_0})\ \phi _m({\vec x_0})$, 
the T-matrix is also separable:
\begin{eqnarray}
\label{5}
\langle \phi _n\ T\ \phi _m\rangle 
=t(\omega )\ \phi _n({\vec x_0})\ \phi _m({\vec x_0}).
\end{eqnarray}
Apart from the trivial solution $\omega=\varepsilon_n$ 
for the case of $\phi _n({\vec x_0})=0$, the poles of $T$ are 
formally given by the roots of the equation 
\begin{eqnarray}
\label{6}
{1 \over {t(\omega )}}= {1 \over v}- G^{(0)}(\omega ) =0
\end{eqnarray}
with
\begin{eqnarray}
\label{7}
G^{(0)}(\omega )\equiv G^{(0)}({\vec x_0},{\vec x_0};\omega )
= \sum\limits_{n=1}^\infty  
{{{\phi _n({\vec x_0})^2} \over {\omega -\varepsilon _n}}}.
\end{eqnarray}
However, eq. (\ref{6}), as it stands, is meaningless since
\begin{eqnarray}
\label{8}
\sum\limits_{n=1}^\infty  
{{{\phi _n({\vec x_0})^2} \over {\omega -\varepsilon _n}}}
&\approx& \ \langle \phi ({\vec x_0})^2\rangle \sum\limits_{n=1}^\infty  
{{1 \over {\omega -\varepsilon _n}}} \nonumber \\
&\approx& \ \langle \phi ({\vec x_0})^2\rangle \ \rho _0\int_0^\infty  
{d\varepsilon {1 \over {\omega -\varepsilon }}}
\to \infty 
\end{eqnarray}
where $\langle \phi ({\vec x_0})^2\rangle$ is the average value of 
$\phi _n({\vec x_0})^2$ among various $n$.  
The divergence is brought about because the density of 
states is constant (which we denote $\rho_0$) 
with respect to the energy.  
To handle the divergence, a scheme for regularization and 
renormalization is called for.  The most mathematically 
satisfying scheme is given by the self-adjoint extension 
theory of functional analysis \cite{AG88,ZO80}.  Here we just quote 
the result.  After the self-adjoint extension, the transition 
matrix $t(\omega)$ is given by
\begin{eqnarray}
\label{9}
{1 \over {t(\omega )}} = \ {{(\omega-i\Lambda )} 
\over {1-e^{i\Theta }}}
\int {d{\vec x}G^{(0)}({\vec x},{\vec x_0};\omega )
G^{(0)}({\vec x},{\vec x_0};i\Lambda )} \nonumber \\
 +{{(\omega+i\Lambda )} \over {1-e^{-i\Theta }}}
\int {d{\vec x}G^{(0)}({\vec x},{\vec x_0};\omega )
G^{(0)}({\vec x},{\vec x_0};-i\Lambda )}.
\end{eqnarray}
Here, $\Lambda$ is an arbitrary scale of the regularization, and
$\Theta$ (0 $\leq$ $\Theta$ $<$ 2$\pi$) is the parameter of 
self-adjoint extension.  
With the straightforward calculation, we find that the energy 
eigenvalues of the system -- the poles of $t(\omega)$ -- are
determined
by the equation
\begin{eqnarray}
\label{10}
{1 \over {\bar v}}-\ \overline G(\omega )\ =\ 0
\end{eqnarray}
where
\begin{eqnarray}
\label{11}
\overline G(\omega )\ =\ \sum\limits_{n=1}^\infty 
{\phi _n({\vec x_0})^2\ 
\left[ {{1 \over {\omega -\varepsilon _n}}+{{\varepsilon _n}
\over {\varepsilon _n^2+\Lambda ^2}}} \right]}
\end{eqnarray}
is the regularized version of $G^{(0)}(\omega)$, and 
\begin{eqnarray}
\label{12}
\bar v=\left[ {{{\Lambda \ \sin \Theta } \over {1-\cos \Theta }}
\sum\limits_{n=1}^\infty  {{{\phi _{n\ }({\vec x_0})^2} 
\over {\varepsilon _{n\ }^2+\Lambda ^2}}}} \right]^{\ -1}
\end{eqnarray}
is the formal (or renormalized) coupling strength of the scatterer.  
We stress  that in spite of purely mathematical construction of 
eqs. (\ref{10})-(\ref{12}), it does correspond to the physical 
small-size limit of the problem of a finite-size obstacle.

        Since the series of eq. (\ref{11}) is convergent, the problem
is now well defined.  We look at the behavior of eq. (\ref{10})
at high energy region $\omega >>\Lambda$.  
For a given value of $\omega$, we can approximate 
eq. (\ref{11}) by truncating the summation at $n=n_x(\omega)$
\begin{eqnarray}
\label{13}
\overline G(\omega )\ 
\approx \ \sum\limits_{n=1}^{n_x(\omega )} 
{\phi _n({\vec x_0})^2\ 
\left[ {{1 \over {\omega -\varepsilon _n}}+{{\varepsilon _n} 
\over {\varepsilon _n^2+\Lambda ^2}}} \right]}
\end{eqnarray}
with an error given by 
\begin{eqnarray}
\label{14}
\delta \overline G\ & = & \ \sum\limits_{n=n_x(\omega )+1}^\infty  
{\phi _n({\vec x_0})^2\ 
\left[ {{1 \over {\omega -\varepsilon _n}}+{{\varepsilon _n} 
\over {\varepsilon _n^2+\Lambda ^2}}} \right]} \nonumber \\
& \approx & \ \langle \phi ({\vec x_0})^2\rangle 
\rho _0\int_{\varepsilon _x(\omega )}^\infty  
{d\varepsilon \left[ {{1 \over {\omega -\varepsilon }}+
{\varepsilon  \over {\varepsilon ^2+\Lambda ^2}}} \right]}
\nonumber \\
& \approx & \ -\langle \phi ({\vec x_0})^2\rangle \ 
\rho _0^2\ {{\ \omega } \over {n_x(\omega )}}
\end{eqnarray}
where we have used 
$\varepsilon _x(\omega )={n_x(\omega)}/{\rho _0}$.
Therefore, we can set
\begin{eqnarray}
\label{15}
n_x(\omega )=\alpha \omega
\end{eqnarray}
where $\alpha$ is a constant inversely proportional 
to the allowable error 
$\delta \overline G$.  Once the summation is truncated at finite 
terms, we can rewrite eq. (\ref{10}) as 
\begin{eqnarray}
\label{16}
{1 \over {v_{e\!f\!\!f}(\omega,{\bar v} )}}\ 
-\sum\limits_{n=1}^{n_x(\omega )} 
{{{\phi _n({\vec x_0})^2} \over {\omega -\varepsilon _n}}}\ =0
\end{eqnarray}
with the effective strength $v_{e\!f\!\!f}(\omega,{\bar v} )$ 
defined through
\begin{eqnarray}
\label{17}
{1 \over {v_{e\!f\!\!f}(\omega,{\bar v} )}}={1 \over {\bar v}}-
\sum\limits_{n=1}^{n_x(\omega )} 
{\phi _n({\vec x_0})^2{{\varepsilon _n} 
\over {\varepsilon _n^2+\Lambda ^2}}}.
\end{eqnarray}
Comparing eq.(\ref{16}) and eq. (\ref{6}), one realizes that 
the problem is now turned into an eigenvalue problem with finite 
basis states $\phi_n$, $n=1,\cdots,n_x(\omega)$.  
Although the system originally has no inherent 
scale, the effective coupling strength $v_{e\!f\!\!f}$ has 
an energy scale $\Lambda$, and, as a result, 
it acquires energy dependence.  
This is possibly the simplest example of the scale anomaly \cite{JA95}.  
The effective coupling $v_{e\!f\!\!f}$ is also referred to as 
{\em running coupling} because of its energy dependence. 
When $v_{e\!f\!\!f}$ is large for the energy region 
of the interest, it induces 
the mixing among the basis states, and results in the Wigner form 
of the nearest neighbor spacing distribution  
$P(s)={1 \over 2}\pi s \exp (-{1 \over 4}\pi s^2)$.
It is known that $v_{e\!f\!\!f}$ is large only in one energy region
determined by the value of $\bar v$ and $\Lambda$ \cite{SH94}.   
Replacing the summation in eq. (\ref{17}) with the integral as 
in eq. (\ref{14}), we have  
\begin{eqnarray}
\label{18}
v_{e\!f\!\!f}(\omega,{\bar v} )\approx {{\bar v} \over
 {1-\bar v\langle \phi ({\vec x_0})^2\rangle
\rho _0 \log \sqrt {1+(n_x(\omega )/ \rho _0\Lambda )^2}}}.
\end{eqnarray}
At the limit $\omega\to\infty$, we have 
$\log \sqrt {1+\ (n_x(\omega )/ \rho _0\Lambda )^2}$
$\approx \log n_x(\omega )$
$\approx \log \omega $.  We arrive at
\begin{eqnarray}
\label{19}
v_{e\!f\!\!f}(\omega,{\bar v} )
\approx -{1 \over
 {\langle \phi ({\vec x_0})^2\rangle \rho _0 \log \omega }}
\ \ \ \ (\omega \to \infty ).
\end{eqnarray}
Remarkably, the formal strength  disappears from the expression
of the effective  strength $v_{e\!f\!\!f}$ in the high energy limit.  
It is now easy to see the reason for the level statistics of 
pseudointegrable billiards becoming more Poisson-like at higher 
energy region,
irrespective to the choice of the formal coupling ${\bar v}$.   
The fact that the strength $v_{e\!f\!\!f}$ disappears at  the limit
$\omega\to\infty$ goes along  well with our intuition that, at the
classical limit, a pointlike obstacle has no effect on the motion of
a particle.   At this limit, all the wave functions are unperturbed 
and the scale invariance is restored in a trivial manner.  
We note that when we replace the sum with the integral 
in obtaining eq. (\ref{18}), we implicitly assume that the 
size of the billiard is far larger than the scale in discussion.  
That is the reason why the size of the billiard boundary, which 
obviously breaks  the scale invariance, does not appear 
in our arguments.

        We next consider the case of two pointlike scatterers.  
We place the scatterers at ${\vec x_0}$ and ${\vec x_1}$ 
with the formal strengths ${\bar v_0}$ and ${\bar v_1}$, respectively.  
The eigenvalues of this system are determined by 
\begin{eqnarray}
\label{20}
\left|
\vphantom{\vrule height5.5ex depth4.0ex}
{\ \matrix
{{\displaystyle{1 \over {\bar v_0}}-\overline G_{00}(\omega)}
&{-G_{01}(\omega )}\cr {-G_{10}(\omega )}
&{\displaystyle{{1 \over {\bar v_1}}}-\overline G_{11}(\omega )}\cr
}\ } 
\right|\ =\ 0
\end{eqnarray}
where $\overline G_{ij}(\omega)$ and $G_{ij}(\omega)$ are defined as
\begin{eqnarray}
\label{21}
\overline G_{ij}(\omega )\ =\ \sum\limits_{n=1}^\infty  
{\phi _n({\vec x_i})\phi _n({\vec x_j})\ 
\left[ {{1 \over {\omega -\varepsilon _n}}
+{{\varepsilon _n} \over {\varepsilon _n^2+\Lambda ^2}}} \right]}
\end{eqnarray}
and
\begin{eqnarray}
\label{22}
G_{ij}(\omega )\ 
=\ \sum\limits_{n=1}^\infty  
{{{\phi _n({\vec x_i})\phi _n({\vec x_j})} 
\over {\omega -\varepsilon _n}}}.
\end{eqnarray}
Let us suppose that two scatterers are placed closely to
each other.
We truncate the sum of $\overline G_{00}(\omega )$ and 
$\overline G_{11}(\omega )$
at $n = n_x(\omega )$ as before.  
As for $G_{01}(\omega )$, which is finite 
as long as ${\vec x_0} \neq {\vec x_1}$, 
the same truncation is possible if two scatterers are apart by
\begin{eqnarray}
\label{23}
\left| {{\vec x_1}-{\vec x_0}} \right|
\simeq {1 \over \sqrt {\varepsilon _x}}
= \sqrt {{\rho _0} \over {n_x(\omega )}},
\end{eqnarray}
since the contributions from higher $n$, which probe
finer length scale than $\left| {{\vec x_1}-{\vec x_0}} \right|$,
cancel among themselves.  Bellow $n \leq n_x (\omega)$,
we can approximate $\phi _n({\vec x_0}) \simeq \phi _n({\vec x_1})$, 
since $\phi_n({\vec x})$ is slowly oscillating in the distance 
$\left| {{\vec x_1}-{\vec x_0}} \right|$. 
The matrix equation, eq. (\ref{20}), then can be reduced to  
\begin{eqnarray}
\label{24}
{1 \over {v_{e\!f\!\!f}^{(2)}(\omega,{\bar v_0},{\bar v_1} )}}\ 
-\sum\limits_{n=1}^{n_x(\omega )} 
{{{\phi _n({\vec x_0})^2} \over {\omega -\varepsilon _n}}}\ =0
\end{eqnarray}
with the effective coupling  
$v_{e\!f\!\!f}^{(2)}(\omega,{\bar v_0},{\bar v_1} )$ given by 
\begin{eqnarray}
\label{25}
v_{e\!f\!\!f}^{(2)}(\omega,{\bar v_0},{\bar v_1} )
=v_{e\!f\!\!f}(\omega,{\bar v_0}) 
+v_{e\!f\!\!f}(\omega,{\bar v_1}). 
\end{eqnarray}
This equation reveals an interesting feature of the system 
with two pointlike scatterers. 
If ${\bar v_0}$ differs from ${\bar v_1}$ appreciably, 
$v_{e\!f\!\!f}(\omega,{\bar v_0})$ and 
$v_{e\!f\!\!f}(\omega,{\bar v_1})$ 
become large at different energies. This means that 
the particle moving in the billiard {\em cannot see} the two 
scatterers at the same time for any given energy.
On the other hand, the two scatterers disturb the particle in a
coherent manner when ${\bar v_0} \simeq {\bar v_1}$. 
With eq.(\ref{19}), we obtain in the limit of 
${\bar v_1}= {\bar v_0}$
\begin{eqnarray}
\label{26}
v_{e\!f\!\!f}^{(2)}(\omega,{\bar v_0},{\bar v_0} )
& = & 2 v_{e\!f\!\!f}(\omega,{\bar v_0} ) \nonumber \\
& \approx & -{2 \over {\langle \phi (x_0)^2\rangle 
\rho _0 \log \omega }}
\ \ \ \ (\omega \to \infty ).
\end{eqnarray}
Namely, two closely placed pointlike scatterers 
with the same formal strength act as a single 
scatterer of twice the {\em effective} strength.  

        If we remove our assumption of two pointlike 
scatterers located closely, we have to deal with eq.(\ref{20}) 
directly.  However, we do not expect an essential change of 
the character of level statistics, since the statistical 
measures are known to be rather insensitive to the precise 
location of the scatterers. This allows us to generalize 
our findings to the case of more than two scatterers. 
Let us consider a billiard with an arbitrary (finite) 
number of pointlike scatterers inside. 
We classify the scatterers according to the magnitude 
of the formal strength; we collect the scatterers 
with the same order of magnitude of the formal strength 
as a single group. We can expect that the scatterers 
belonging to one of such groups disturb the particle motion 
in a coherent manner in the energy region determined by 
their own formal strength, while their influence never appears 
at different energies. In particular, in the case of scatterers 
with a common formal strength, their effects are additive.
This essentially explains the known behavior of  pseudointegrable
billiards that by increasing the number of singular points one
obtains more Wigner-like level statistics. 

        We now place our findings in a broader context.  
The standard approach in quantum chaology has been the 
semiclassical periodic orbit theory \cite{GU70,BO89} which is a 
truncation of WKB approximation.  It cannot be easily 
applied to the problem with singular scatterers, since the 
classical orbit hitting the pointlike scatterer occupies 
a set of measure zero in the phase space.  Our approach, 
on the contrary, is fully quantum mechanical without any 
resort to semiclassical approximations.  Although the present 
approach is applicable only to the system with pointlike 
scatterers, our results certainly give an insight to the 
problem of quantum level statistics of generic 
pseudointegrable system.  Specifically, we believe that it 
establishes a firm ground to the earlier intuitive notion 
of "wave chaos" \cite{CC89,SE90}, that is the chaotic motion 
generated by the uncertainty principle in the motion of 
wave-natured quantum particles.

        The system studied here clearly is a pedagogical 
example of the quantum violation of classical scale invariance.  
We believe that it enhances the intuitive understanding of the 
scale invariance and asymptotic freedom which have been viewed 
as phenomena found only in the esoteric theories of elementary 
particles.  Finally, we stress that, even apart from the 
arguments on quantum chaos, we could have arrived at the level 
statistics in search of the measurable consequence of scale 
anomaly of the system, since it gives a handy way to see the 
effective strength of the pointlike scatterer in the absence 
of clear-cut measures to gauge its effect on the individual states.

        In summary, we have derived the energy and number-of-scatterer 
dependence of the effective coupling strength of pointlike scatterers 
in billiard systems.  We have shown that it explains the key 
characteristics of the level statistics of the quantum 
pseudointegrable billiards.

        We express our thank to Profs. Izumi Tsutsui
and Hidezumi Terazawa for the 
helpful discussions and comments.  TC gratefully acknowledge 
the hospitable research environment offered to him by the members of the
Theory Division at Institute for Nuclear Study, University of Tokyo.  
TS acknowledge the support of the Grant-in-Aid for Encouragement 
of Young Scientists (No.07740316) by the Ministry of Education, 
Science, Sports and Culture of Japan.
\end{document}